\documentclass[aps,pre,twocolumn,superscriptaddress,showpacs]{revtex4-1}

\usepackage{amsmath,amssymb,bm}
\usepackage{epsfig}
\usepackage{graphicx}
\usepackage{graphics}
\usepackage{xcolor}

\begin{document}

\title{Emergence of oscillations in fixed energy sandpile models on complex networks}

\author{Davood Fazli, Nahid Azimi-Tafreshi\\Physics Department, Institute for Advanced Studies in Basic Sciences, Zanjan 45137-66731, Iran}

%
%

\begin{abstract}
Fixed-energy sandpile (FES) models, introduced to understand the self-organized criticality, show a continuous phase transition between absorbing and active phases. In this work, we study the dynamics of the deterministic FES models on random networks. We observe that close to absorbing transition the density of active nodes oscillates and nodes topple in synchrony. The deterministic toppling rule and the small-world property of random networks lead to the emergence of sustained oscillations. The amplitude of oscillations becomes larger with increasing the value of network randomness. The bifurcation diagram for the density of active nodes is obtained.
We use the activity-dependent rewiring rule and show that the interplay between the network structure and the FES dynamics leads to the emergence of a bistable region with a first-order transition between the absorbing and active states. Furthermore during the rewiring, the ordered activation pattern of the nodes is broken, which causes the oscillations to disappear.
\end{abstract}


\maketitle

\section{Introduction}

The idea of self-organized criticality (SOC), introduced by Bak-Tang-Weinberg (BTW), is one of the most major concepts for understanding the physics of complex systems \cite{Bak}. A system displaying SOC organizes itself into a critical state with the statistical properties governed by power laws \cite{Tang, wat}. Unlike the standard critical systems, approaching the critical point occurs without tuning a control parameter. Avalanches with power law distribution emerge as a result of external driving and dissipation.

The BTW sandpile model is the simplest well-studied model which displays SOC behavior \cite{Bak, Dhar, Dhar2}. The model is simply defined on a square lattice. To each site, an integer height variable is assigned which represents the number of the sand grains of that site. A configuration in which all sites have heights less than a specific threshold is stable. The dynamics of the model is started from a stable configuration. At each step a grain of sand is added to a randomly chosen site so that its height is increased by $1$. If the height of that site reaches its threshold value, that node becomes unstable, topples and loses some grains of sand, redistributed among its nearest neighbors. As a result of the toppling, some of the neighbors may also become unstable. The toppling process continues until all sites become stable and an avalanche is formed. Avalanches that reach the boundaries allow sand to exit the boundaries. The balance between the external driving and dissipation tunes the system to the stationary state with  critical properties.

\begin{figure*}[t]
\begin{center}
\scalebox{0.63}{\includegraphics[angle=0]{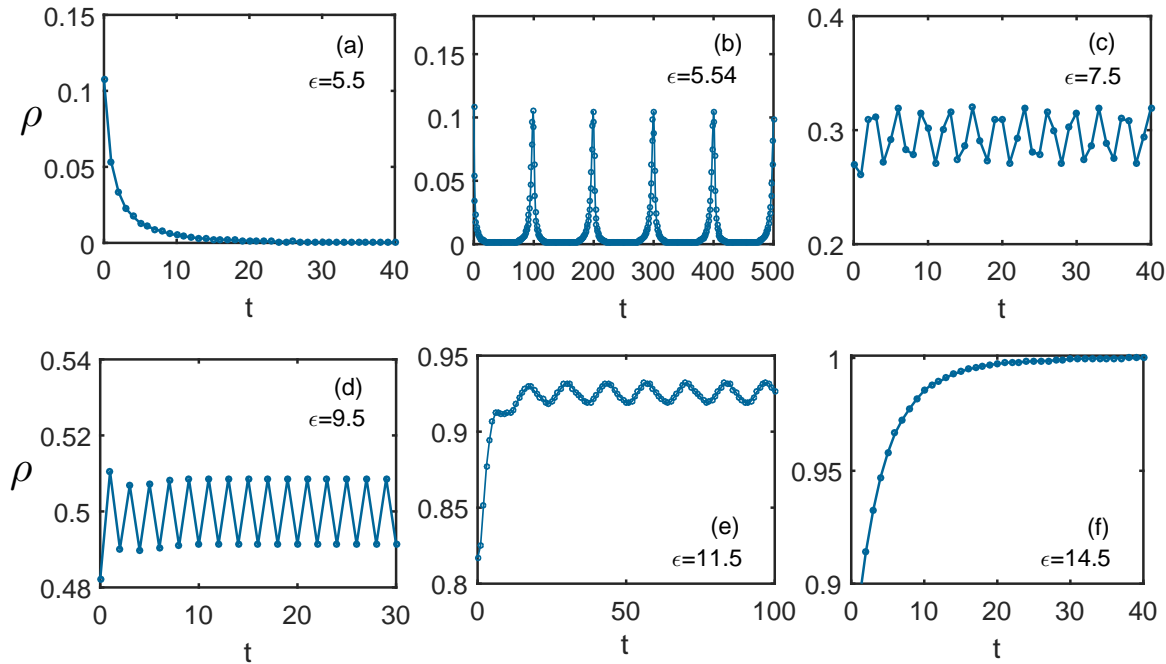}}
\end{center}
\caption{Time evolution of the density of active nodes, $\rho(t)$, for (a) $\epsilon < \epsilon_{c1}$, (b), (c) $\epsilon_{c1} <\epsilon < \epsilon_{c2}$, (d) $\epsilon_{c2} <\epsilon < \epsilon_{c3}$, (e) $\epsilon_{c3} <\epsilon < \epsilon_{c4}$ and (f) $\epsilon > \epsilon_{c4}$, on an ER network with $N=10^4$ and $\langle k\rangle=10$.}
 \label{f1}
\end{figure*}

To explain how the sandpile model self-organizes into a critical state, Dickman \textit{et al.} related the standard sandpile model to a conservative system, the so-called fixed energy sandpile (FES) model \cite{Dikman, FES, FES2}. In this model driving and dissipation are set to zero, such that the total number of sands (which can be interpreted as energy) is constant. A site at which the number of grains is equal to or greater than the threshold value is called active. The dynamics of the model is started with a number of initial active sites. At each time step, the active sites topple and redistribute the sands to their neighbors and a new configuration with a set of active and inactive nodes is created. The process reaches either an absorbing state, in which no active site exists, or an active state with a constant fraction of active sites. In the deterministic FES models, the activity configurations are revisited after a period of time. Since the system is closed, for any finite system size the number of configurations is finite and the system enters a periodic cycle after a transient time \cite{Bagnoli, Lucca}.

The fixed energy sandpile model exhibits a second order phase transition, as the energy density grows, from an absorbing to an active phase \cite{PT1, PT2, montakhab, path}.
The SOC tunes the system to the edge of this transition through slow driving and dissipation. Hence, the scaling properties of the SOC system can be found from the critical properties of the absorbing phase transition in the FES model \cite{FES}. By extending the SOC to systems exhibiting bistability and phase coexistence, the concept of self-organized bistability (SOB) has been recently proposed as a mechanism through which a system self-organizes to the edge of bistability of a discontinuous phase transition \cite{munoz1, munoz2}.

In general, the BTW sandpile dynamics can be defined on a random network with any arbitrary degree distribution \cite{Bonabeau, Goh1}. In this case, the threshold value for each node is different and is assumed to be equal to the degree of each node. The dissipation of sand grains occurs either through some sinks from which grains flow out of the system, or by assuming that at each time step a fraction of the grains is lost during the toppling. It is found that avalanche power-law distributions and critical exponents depend on the degree distribution and structure properties of the underlying network \cite{Goh2}.
\begin{figure}[t]
\begin{center}
\scalebox{0.42}{\includegraphics[angle=0]{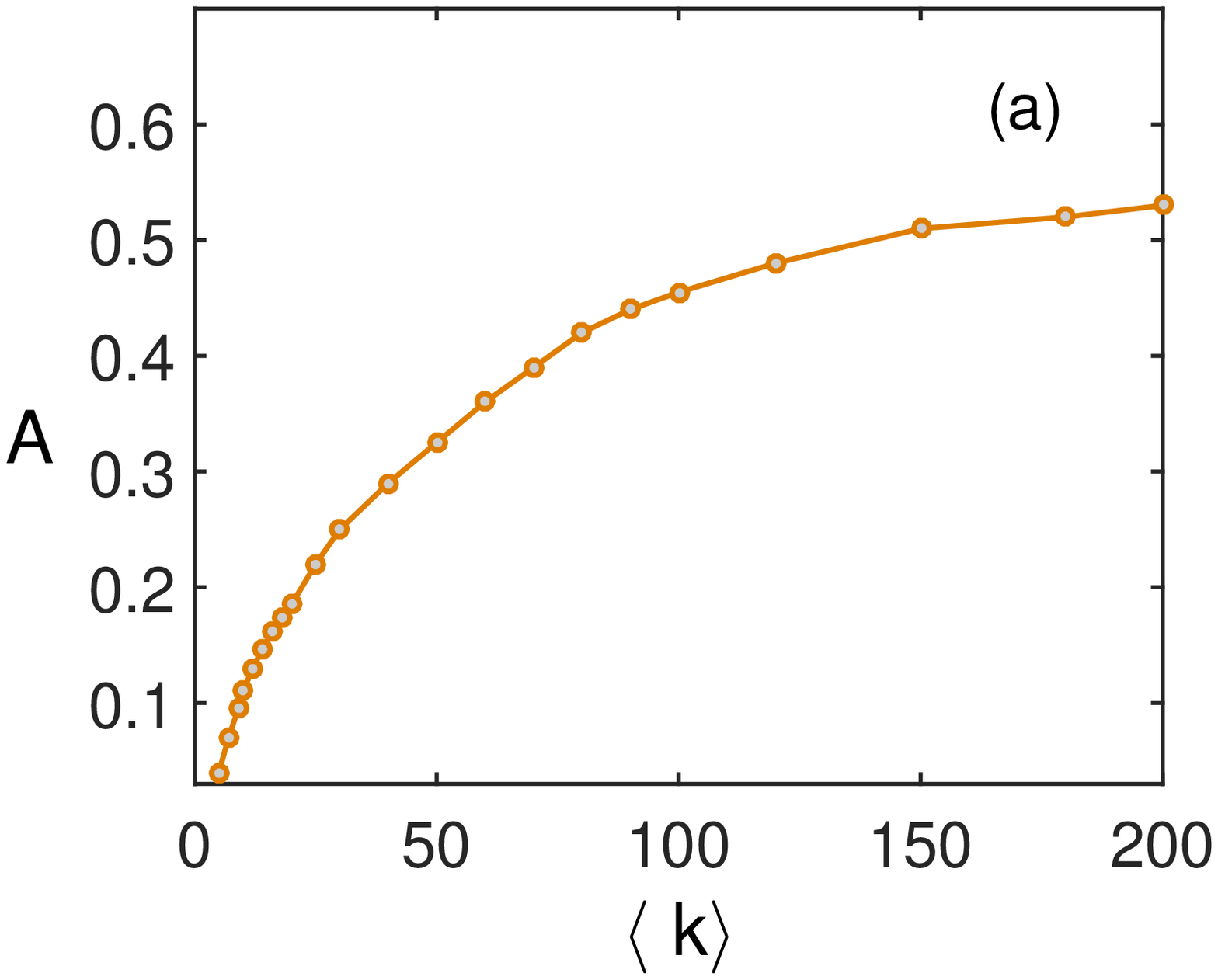}}
\scalebox{0.42}{\includegraphics[angle=0]{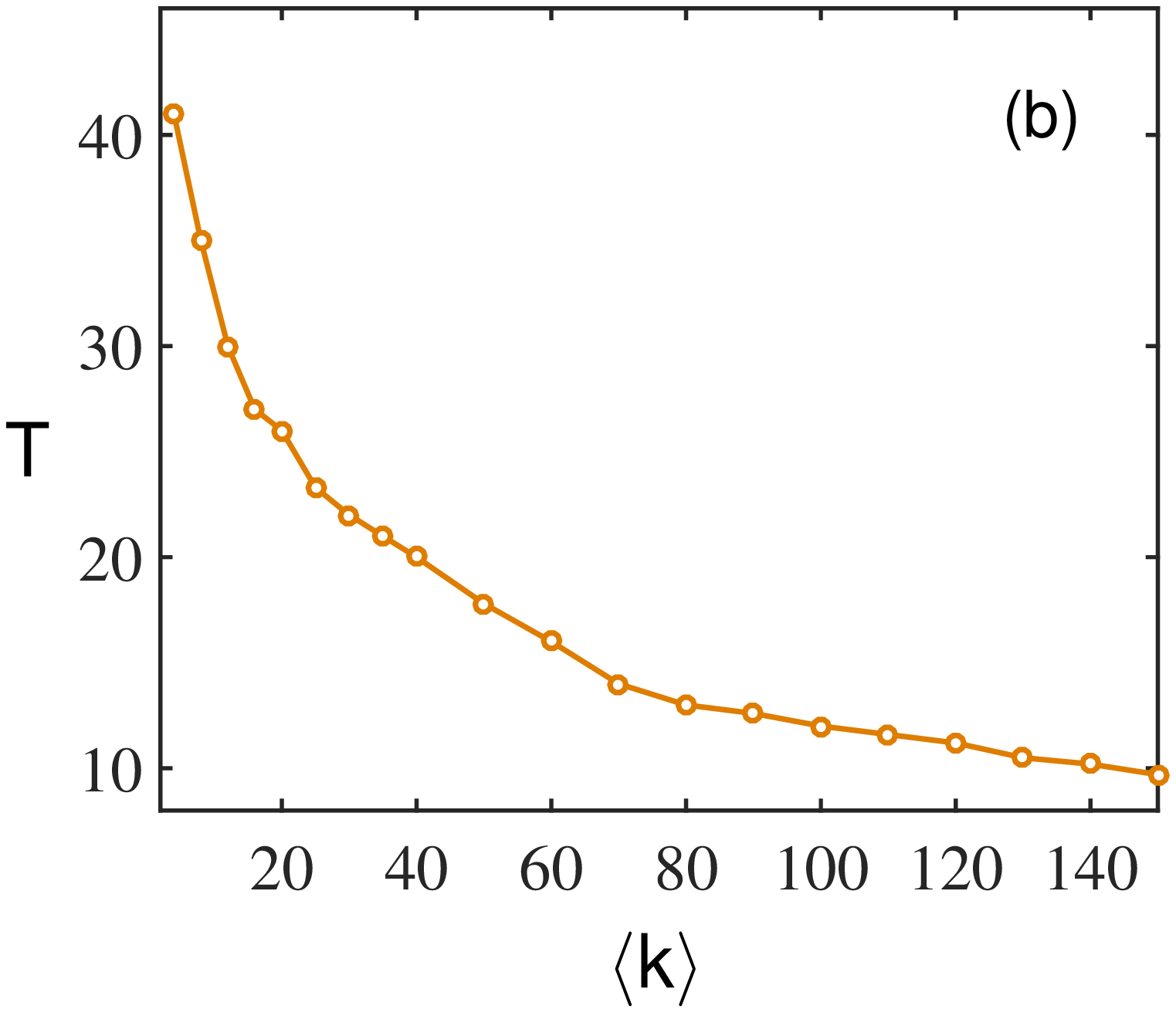}}
\end{center}
\caption{$(a)$ Amplitude and $(b)$ time period of the oscillations at the transition point $\epsilon_{c1}$ for each mean degree value, averaged over $10^3$ trails.}
\label{f11}
\end{figure}
\begin{figure}[t]
\scalebox{0.45}{\includegraphics[angle=0]{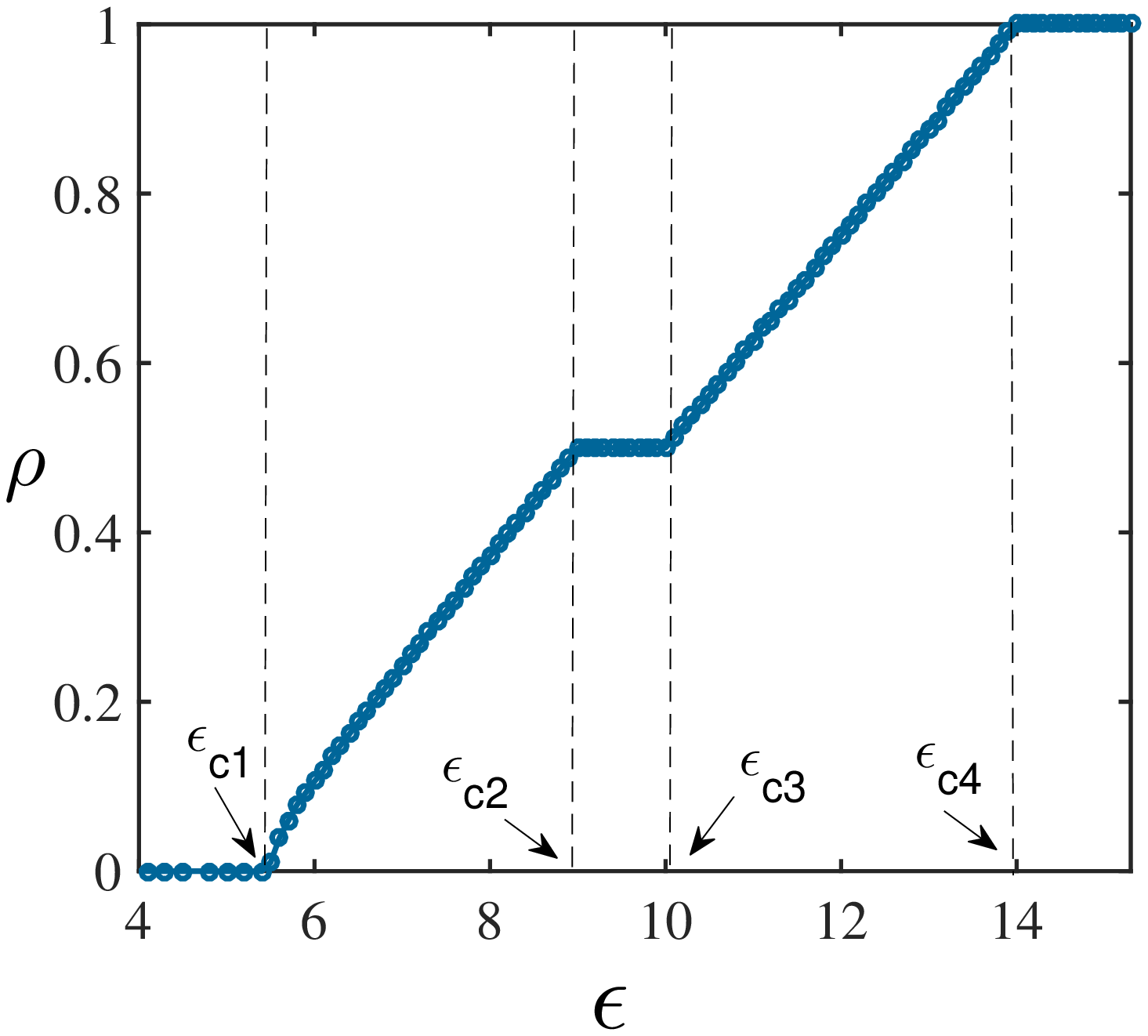}}
\caption{Stationary values of the density of active nodes as a function of $\epsilon$, on an ER network with $N=10^4$ and $\langle k \rangle=10$, averaged over $100$ trails. Transition points occur at $\epsilon_{c1}=5.51$, $\epsilon_{c2}=8.95$ , $\epsilon_{c3}=10.05$ and $\epsilon_{c4}=13.96$.}
\label{f2}
\end{figure}

As the structure of a network changes the avalanches distribution of the sandpile model, the sandpile dynamics can affect the network structure. The coevolution of the network structure and dynamics of the processes taking place in the network is studied in the framework of adaptive networks \cite{adaptive1, adaptive2}. Bianconi \textit{et al.} \cite{Bianconi} proposed a simple model of network dynamics, which has the same dynamics as the sandpile model. The dynamics occurs in two parts: slow driving by adding new edges to the network and fast relaxation dynamics of avalanche of rewiring processes with dissipation. They showed that the network consequently self-organizes to a critical point in which avalanches present scaling behaviors. Also, in Ref. \cite{coevolution}, it was shown how such an interplay between the sandpile dynamics and network structure leads to self-organization to a critical point in which the distribution of avalanches and degree distribution of the network similarly behave as power-law functions.

In this paper we consider the FES model with the BTW toppling rule (BTW-FES) on random networks. It is interesting to see the effect of the small-world property on the dynamics of this model. Using numerical simulations, we show that close to absorbing phase transition, the active sites topple in synchrony and the density of active sites oscillates. With increasing the network randomness, the amplitude of sustained oscillations grows. We find that a transition to synchronization occurs as a function of the network randomness. For regular lattices it was shown that with increasing the energy density, the density of active nodes increases as a devil’s staircase \cite{Bagnoli}. We find that the number of constant plateaus decreases as the value of network randomness increases and finally for a fully random network there is only one plateau with a $2-$period cycle.

In the following we use the interplay of structural effects and the BTW-FES dynamics. We rewire the network edges due to the state of nodes. As the result of rewiring, a region of bistability emerges in which both the absorbing state and the active state are stable. Furthermore rewiring destroys the ordered pattern of topplings among the nodes which makes the sustained oscillations disappear.

The article is organized as follows. In the next section, we define the dynamics of the BTW-FES on a random network and obtain the phase diagram of the model. In Sec.~III we discuss the role of the small-world effect in the emergence of sustained oscillations in the dynamics of sandpile models.
The interplay between the network structure and the sandpile dynamics is discussed in Sec. IV. The paper is concluded in Sec. V.

\section{The model}

Let us consider an Erd\H{o}s--R\'enyi (ER) random network with $N$ nodes and mean degree $\langle k \rangle$, a paradigmatic example of a homogeneous network. We define the BTW-FES on this network as follows. A conserved quantity $E$ is considered as the total number of sand grains on the network. The quantity $E$ can be also interpreted as the total energy, such that the parameter $\epsilon=E/N$ is the energy density of the system. Each node has a threshold $h^c_{i}$, which is the capacity of the sand or energy that the node can hold. We set the threshold values equal to the degree of each node such that $h^c_{i}=k_i$. At the initial state, the total energy $E$ is divided randomly among $N$ nodes such that each node gets $h_i$ sands or energy. The nodes at which $h_i \geq h^c_i$ are called active, otherwise they are inactive. In other words at the initial state some nodes are active and the rest are inactive. Hence we can consider the model as a binary activation model with states $s_i=0$ for the inactive nodes and $s_i=1$ for the active nodes, such that the density of active nodes is defined as $\rho(t)= \frac{1}{N}\sum_i s_i(t)$ at each time step $t$. The activation dynamics occurs as follows. At each time step, each of the active nodes topples once following the same rule as the standard BTW model, i.e., $h_i= h_i-k_i$ and each of its neighbors gets one sand such that $h_j=h_j+1$. Active nodes can still have sands more than the threshold and therefore remain active. Also inactive nodes can become active as a result of toppings. All active nodes will topple at the next time step and the process continues until the system reaches a stationary state.

We carried out Monte Carlo simulations of the model on an ER network with $N=10^4$ and $\langle k \rangle=10$. Figure~\ref{f1} shows the evolution of the density of active nodes. The stationary state depends on the value of energy density $\epsilon$. For $\epsilon$ less than a critical value $\epsilon_{c1}$, the model shows an ``absorbing" phase, in which $\rho_{stat}=0$ (Fig.~\ref{f1} $(a)$). That is the same as what happens on a lattice. However for $\epsilon > \epsilon_{c1}$, the density of active nodes oscillates in the stationary state.
Close to the critical point $\epsilon_{c1}$, the oscillations are regular but by increasing $\epsilon$ the irregular oscillations appear (Fig.~\ref{f1} $(b),(c)$). A super-critical Hopf bifurcation takes place at $\epsilon_{c1}$
with the appearance of the limit cycle for the density of active nodes and the ``oscillatory" phase is observed in the range $\epsilon_{c1}< \epsilon < \epsilon_{c2}$. In the oscillatory phase, the amplitude $A$, and the time period $T$ of oscillations depend on the mean degree of the network. With increasing $\langle k\rangle$, the time period decreases, while the amplitude of oscillations increases (Fig.~\ref{f11}).

\begin{figure}[t]
\begin{center}
\scalebox{0.45}{\includegraphics[angle=0]{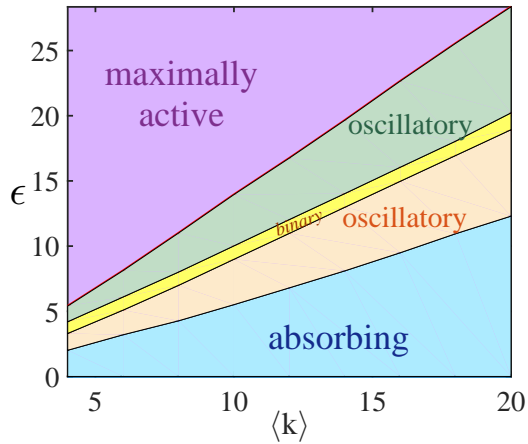}}
\end{center}
\caption{Phase diagram of the BTW-FES model on an ER network with $N=10^4$ in the space $\epsilon-\langle k \rangle $.}
\label{f3}
\end{figure}

At the transition point $\epsilon_{c2}$, we observe that the active nodes start to oscillate between two states, up and down and the time period is $T=2$ (Fig.~\ref{f1} $(d)$). We call this regime, which is in the range $\epsilon_{c2}< \epsilon < \epsilon_{c3}$, the ``binary" state. For $\epsilon > \epsilon_{c3}$ the density of active nodes increases to more than half while the density of inactive nodes decreases. Hence the model shows a symmetrical behavior such that the behavior of the inactive and active node densities is exchanged. Therefore for $\epsilon_{c3}< \epsilon < \epsilon_{c4}$, the oscillations appear again (Fig.~\ref{f1} $(e)$). With further increase $\epsilon$, the density of active nodes goes up while the amplitude of the oscillations decreases. Finally, at the transition point $\epsilon_{c4}$, the oscillations disappear and the density of active nodes starts approaching the trivial fixed point $\rho_{stat}=1$, in which each node is active. We call the region $\epsilon > \epsilon_{c4}$ the ``maximally active" phase (Fig.~\ref{f1} $(f)$). Figure~\ref{f2} shows the density of the active nodes as a function of $\epsilon$. The transition points depend on the initial state. For a uniformly random initial distribution of the sands, the transition points occur at $\epsilon_{c1}=5.51\pm0.01$, $\epsilon_{c2}=8.95\pm0.01$ , $\epsilon_{c3}=10.05\pm0.01$ and $\epsilon_{c4}=13.96\pm0.01$. The activity shows a step-like behavior with only one plateau (binary state) and there is a symmetry around the plateau. We can also see the different phases of the model in Fig.~\ref{f3}. As we can see the transition points also depend on the mean degree of the network and shift to higher values for the networks with more connectivity.
\begin{figure}[t]
\begin{center}
\scalebox{0.58}{\includegraphics[angle=0]{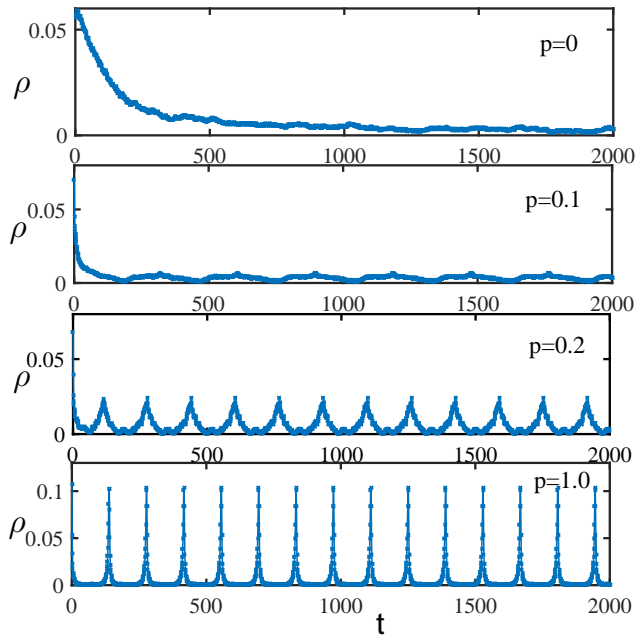}}
\end{center}
\caption{Time evolution of the density of active nodes for $\epsilon=5.7$ on small-world networks with different values of rewiring probability $p$.}
\label{f4}
\end{figure}

\section{Emergence of rhythmic behavior}

In the BTW-FES model on ER random networks the density of active nodes oscillates in each cycle. That is different from the well-known behavior of the model on regular lattices in which the density of active nodes reaches a stationary state fixed point. The root of the oscillations can be in the small-world property of random networks. Let us consider the BTW-FES on a small-world network, produced following the Watts-Strogatz algorithm \cite{watts}. We start with a ring having $N$ nodes and node degree $k=2m$, such that each node is connected to its $m$ nearest neighbors on each side. Moving clockwise, for every node we select randomly an edge that connects that node to one of its neighbors, and rewire it with probability $p$. We continue this process until each edge in the original ring has been considered once. The parameter $p$ measures the degree of disorder or randomness
of the resulting network. For $p = 0$ the network is regular and fully ordered, while for $p=1$ all edges are rewired and the resulting network corresponds with the ER random network.

\begin{figure}[t]
\begin{center}
\scalebox{0.42}{\includegraphics[angle=0]{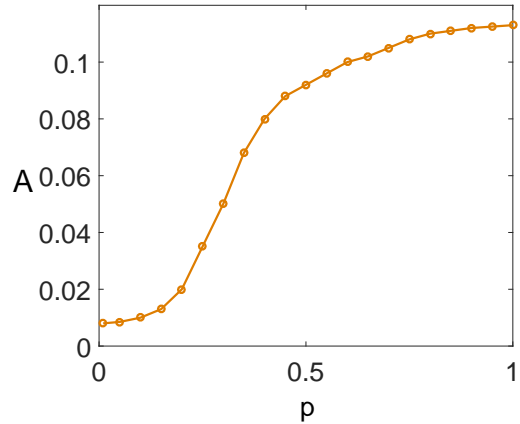}}
\end{center}
\caption{The amplitude of oscillations as a function of the network randomness at $\epsilon=5.7$, averaged over $10^3$ trails.}
\label{f41}
\end{figure}

We consider the dynamics of the BTW-FES on a small-world network with $N=10^4$ and $m=5$. At the initial time the fraction $\epsilon =5.7$ of sands are randomly distributed among the nodes of the ring. Figure \ref{f4} shows the evolution of the density of active nodes for some values of $p$.  As we can see, for $p=0$ (a regular ring), the density of active nodes decreases with the time and reaches a stationary fixed point. However with increasing the value of $p$, oscillations emerge in the stationary state. As we can see in Fig.~\ref{f41} for higher values of $p$, the amplitude of the oscillations is greater.
\begin{figure}[t]
\begin{center}
\scalebox{0.4}{\includegraphics[angle=0]{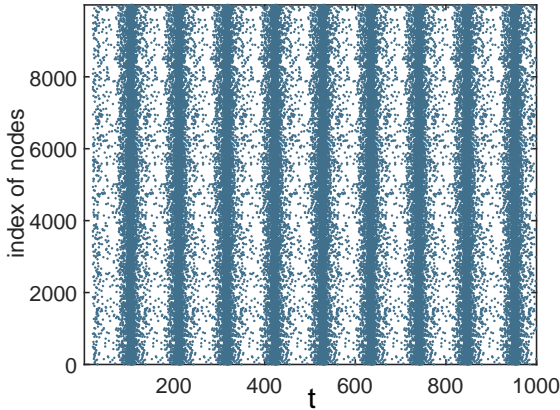}}
\end{center}
\caption{Raster plot for the activation of nodes in a small-world network with $m=5$ and $p=1$. The density of sand is $\epsilon=\epsilon_{c1}=5.51$.}
\label{f5}
\end{figure}
\begin{figure}[t]
\begin{center}
\scalebox{0.4}{\includegraphics[angle=0]{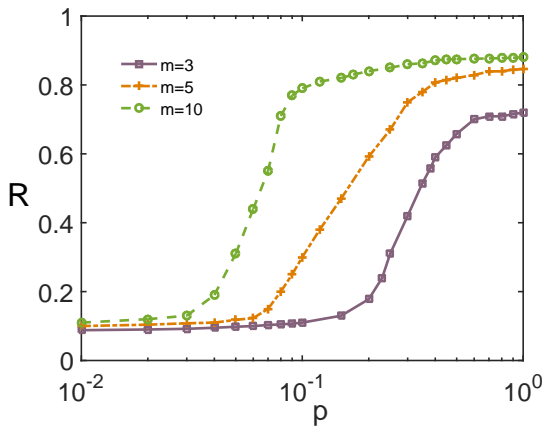}}
\end{center}
\caption{Synchronization order parameter $R$ as a function of $p$ on small-world networks with $N=10^6$. The curves are plotted at $\epsilon_{c1}$ for different mean-degree values $\langle k\rangle = 2m$.}
\label{f6}
\end{figure}
\begin{figure*}[t]
\begin{center}
\scalebox{0.35}{\includegraphics[angle=0]{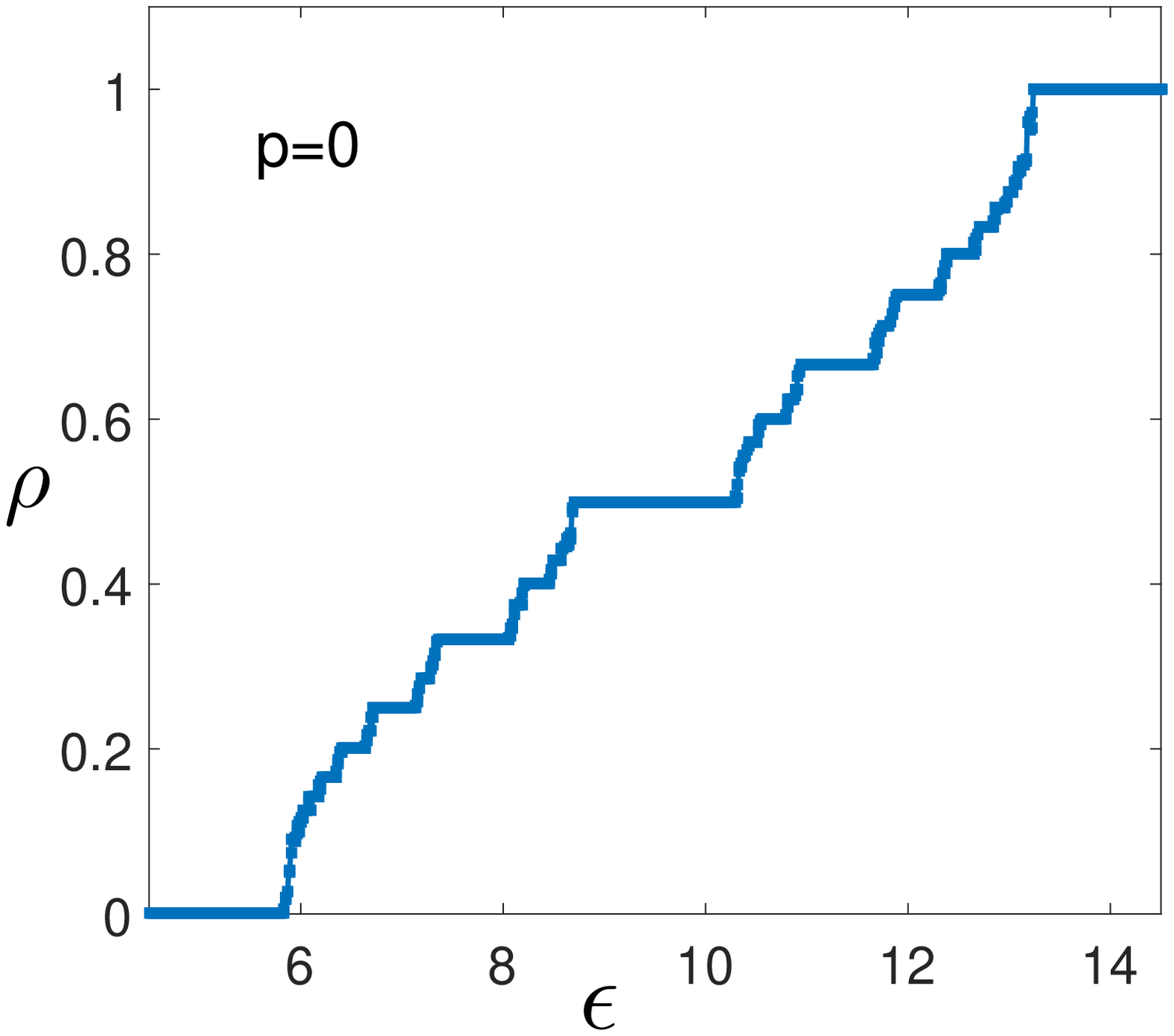}}
\scalebox{0.35}{\includegraphics[angle=0]{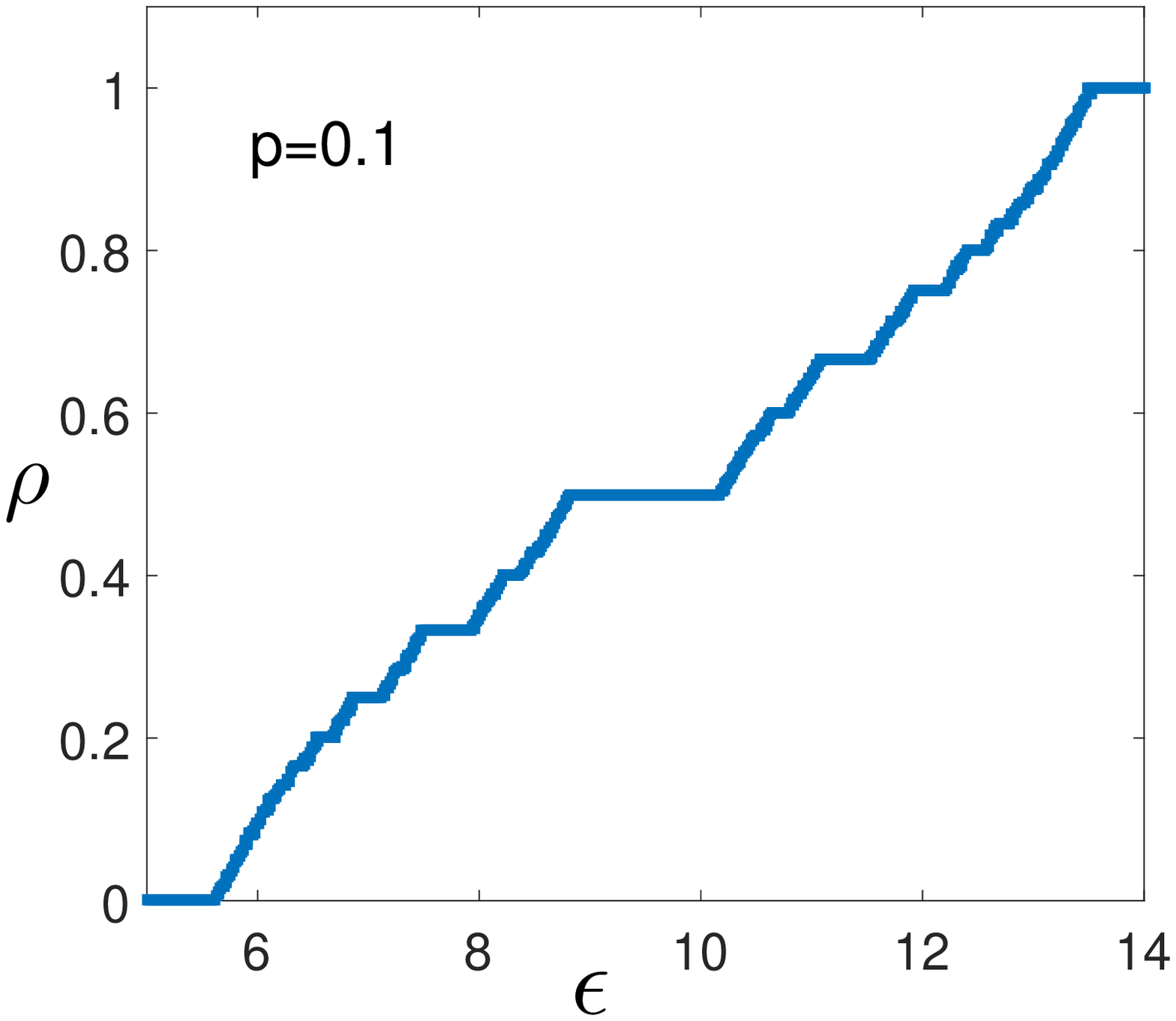}}
\\
\scalebox{0.35}{\includegraphics[angle=0]{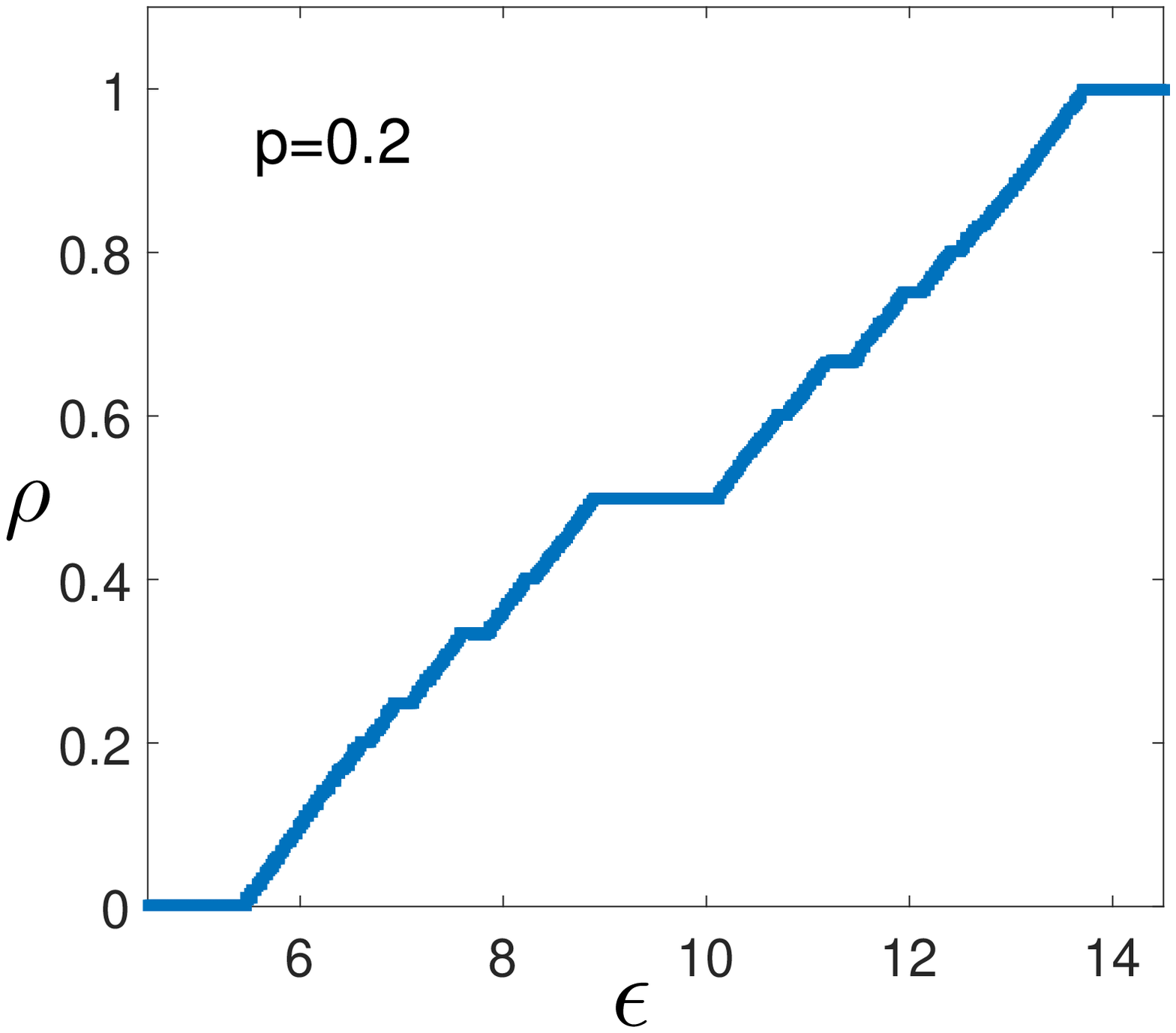}}
\scalebox{0.35}{\includegraphics[angle=0]{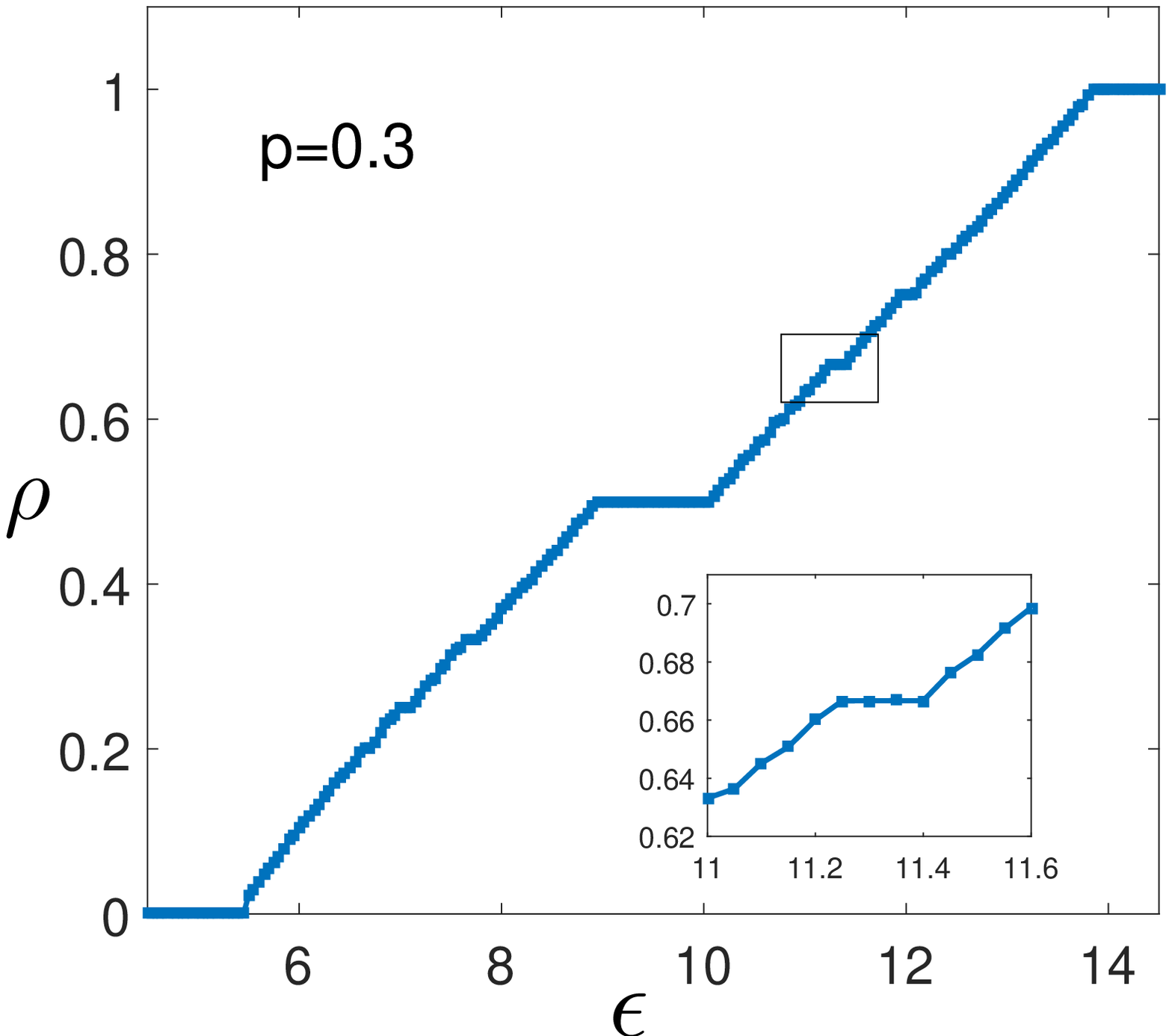}}

\end{center}
\caption{ Stationary values of the density of active nodes as a function of $\epsilon$, on a network with $N=10^4$ and $\langle k\rangle=10$, averaged over $100$ trails, and for different values of network randomness $p$.}
 \label{f61}
\end{figure*}
\begin{figure}[t]
\begin{center}%
\scalebox{0.46}{\includegraphics[angle=0]{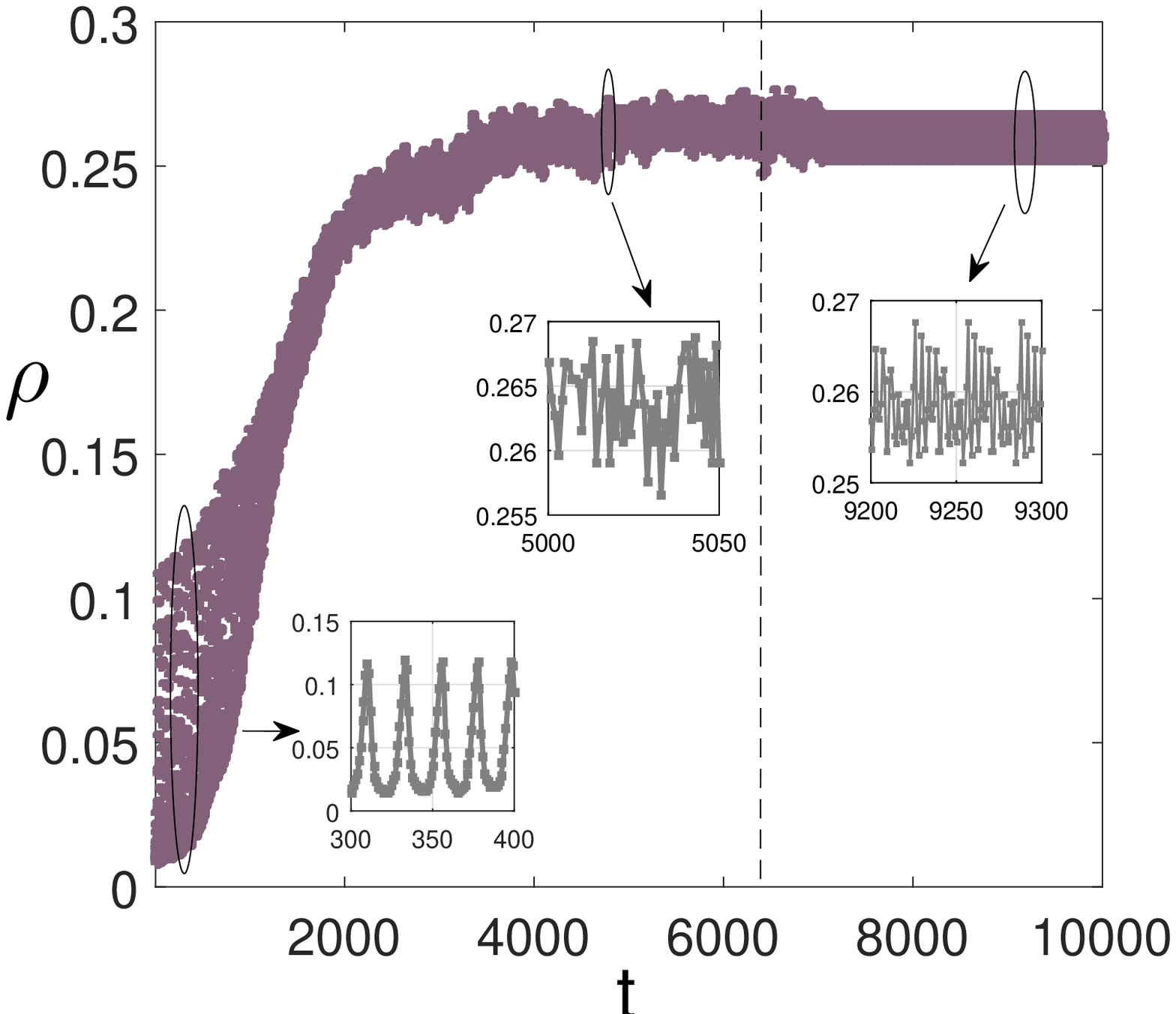}}
\end{center}
\caption{Time evolution of the density of the active nodes for the BTW-FES model on an adaptive network. The initial network is an ER random network with $N=10^4$ and $\langle k\rangle=10$. At each time step edges are rewired with probability $f=0.1$. The dashed line indicates the time at which the rewiring stops.}
\label{f8}
\end{figure}

The oscillations correspond to a synchronized behavior for the active nodes. At each step, any node with index $j$ is active. These active nodes are characterized by a time counter $\tau_j$, which is zero at that step. The active nodes $j$ become inactive altogether at the next step. The time counter $\tau_j$ counts the number of steps that this set of nodes remains inactive and reaches the final value $T$ when the nodes with index $j$ will be active again. We can see this periodic pattern for the activation of the nodes in Fig.~\ref{f5}. Active nodes are represented by blue (dark) points, and white points show the inactive nodes. The density of active nodes oscillates during a periodic cycle. Let us start a cycle with an configuration in which $\rho$ is maximum. The density of active nodes decreases in the next configurations and reaches its minimum in the middle of the cycle. The activity density increases again and reaches the initial maximum value when the initial configuration is revisited and the cycle is completed. The oscillating behavior of the density of active nodes indicates that a significant fraction of nodes activates synchronously. At each time step $t$, we can assign a phase $\phi_j(t)$ to each node $j$, defined as $\phi_j(t)= 2\pi(\tau_j(t)-1) / T$. Hence, the order parameter for synchronization of phases $\phi_j$ at time step $t$ is defined as,
\begin{equation}
R(t) = \big|\frac{1}{N}\sum _j ^N e^{i \phi_j (t)}\big|.
\label{eq1}
\end{equation}

Close to the critical point $\epsilon_{c1}$, a large number of nodes are synchronized and are activated together, such that the value of $R$ is large. The synchronization is complete, $R=1$, if all nodes become active at the same time. However as we move away from the critical point, the nodes start to be  activated irregularly, such that the corresponding phases become different and are distributed between $[0, 2\pi]$. In this case the order parameter $R$ gets close to zero.

Figure.~\ref{f6} shows the behavior of $R$ in the stationary state in terms of the parameter $p$. To see a better synchronized behavior we consider a larger network with $N=10^6$ nodes. Since close to the critical point $\epsilon_{c1}$, the activated nodes are more synchronized, we consider $\epsilon \sim \epsilon_{c1}$ for each network with a definite value of the mean degree. We can see a phase transition occurs from a non-synchronized to a synchronized state at a definite value $p_c$. The value of $p_c$ depends on the number of neighbors $2m$. For a network with larger mean degree, the synchronization emerges at a smaller value of $p_c$.

The network disorder also affects on the devil's staircase behavior of the activity density as a function of energy. For a regular network there are many plateaus. As we can see in Fig.~\ref{f61} with increasing the degree of network randomness (rewiring probability $p$), the number of constant plateaus decreases and finally for a fully random network $(p=1)$, there is only one plateau with a $2$-period cycle. There is a symmetry around the plateau and the activity synchronization reappears approaching the maximally active sate.

\section{Activity-dependent rewiring}

So far, we have considered the dynamics of the fixed energy sandpile models on static networks. In this case the degree of nodes and thus the value of thresholds are constant. Here we allow the network structure changes during the dynamics. We consider a simple rewiring rule such that during the dynamics of the BTW-FES model, active nodes avoid to connect with each other. In this case, the toppling of the active nodes makes it possible for more inactive nodes to be activated. Also the threshold values change during the rewiring and then the ordered pattern for the toppling of the nodes breaks down.

Let us consider the BTW-FES model on the ER random network with $N=10^4$ and $\langle k\rangle=10$. We take $\epsilon=\epsilon_{c1}=5.51$. The sands are randomly distributed among the nodes.
Hence, the initial state of the nodes is $(s_1(0),\ldots s_N(0))$, such that some nodes are active $(s_i(0)=1)$ and the rest are inactive $(s_i(0)=0)$ initially. At each time step, we first apply the following rewiring rule for $f$ fraction of the active edges (edges which connect two active nodes). An active edge is randomly chosen and from one of its active end nodes, it is disconnected and rewired to a randomly chosen inactive node in the network. Then we apply the dynamics of the BTW-FES model and allow all active nodes to topple. At the next time step, we repeat the rewiring rule and synchronously update the state of the nodes. The process is continued until the system reaches a stationary sate.

Figure~\ref{f8} shows the evolution of the active node density during the rewiring process. As we can see, during rewiring the density of active nodes increases. Whenever an active node topples, it loses as much sand as its degree. However, the node degree and therefore the activation threshold of the nodes change due to rewiring such that the ordered activation pattern of the nodes is broken. That is why oscillations become smaller and after a while the dynamics evolves to a stationary fixed point.
At this point, we stop the rewiring and allow the dynamics of the FES model to proceed. After stopping the rewiring, we still have a random network that has the same mean degree as the network before the rewiring. Hence, as we expect after a while the oscillations emerge again due to the small-world effect and the order of topplings among the nodes. However, the rewiring causes that the arrangement of the active and inactive nodes changes and therefore the density of the active nodes oscillates around a higher mean value.

\begin{figure}[t]
\begin{center}
\scalebox{0.42}{\includegraphics[angle=0]{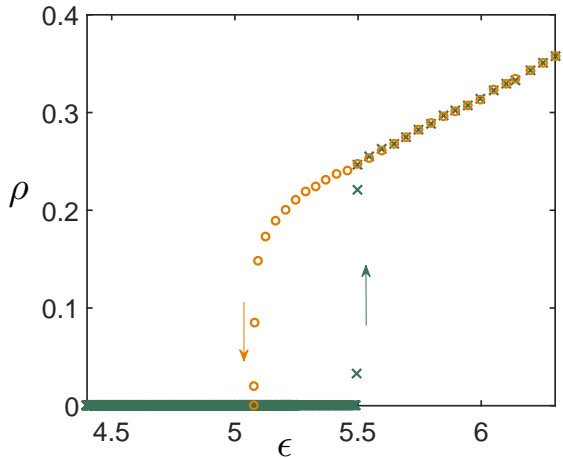}}
\end{center}
\caption{Phase transitions of the BTW-FES model on an adaptive network. The initial network is an ER random network with $N=10^4$ and $\langle k\rangle=10$.}
\label{f9}
\end{figure}

Figure~\ref{f9} shows the steady state density of the active nodes $\rho$
as a function of the density of sand grains $\epsilon$. We start from the absorbing state and increase the sand density $\epsilon$ slowly. At each step we add a small amount $\Delta \epsilon$ to the stationary configuration of the previous step, and distribute $N \Delta \epsilon$ grains randomly among the nodes and let the system reaches a new stationary configuration. In reverse sweep, similarly at each step we remove $N \Delta \epsilon$ grains from random nodes of the previous configuration.

As we already saw, in the case of static networks a continuous absorbing phase transition occurs. However by including rewiring, a bistable region is observed in which the active and absorbing phases are both stable. The coevolution of structure and dynamics makes the system abruptly jumps from an absorbing phase to an active phase. On the other hand with decreasing of sand density, the system tends to remain in the active phase and finally collapses in the absorbing phase in a lower value of $\epsilon$. Thus, a hysteresis loop is formed. This observation can be related to the concept of self-organized-bistability, such that the system self-organizes to a bistable region instead of a critical point.

\section{Conclusion}

In this work, we studied the dynamics of the deterministic BTW-FES model using numerical simulations on random networks. The model undergoes an absorbing phase transition from a fixed point with zero density of active nodes to an oscillatory active phase. The density of active nodes oscillates during a periodic cycle. We discussed that the small-world effect and symmetric redistribution of sand grains lead to the emergence of sustained oscillations. For higher values of network randomness the amplitude of the oscillations increases. By increasing the sand density, the model undergoes other subsequent phase transitions and binary phase and active fixed point emerge. We obtained the phase diagram of the model and found the transition points for random networks with different mean degrees. The values of transition points depend on the mean degree of the network and the initial distribution of sand (energy).

The ordered toppling pattern can be destroyed by rewiring the network edges during the BTW-FES dynamics. In the rewiring process, we assumed that the active nodes avoid to connect with each other. In this case, the density of active nodes grows up and the oscillations will gradually disappear during the rewiring. Moreover, the coevolution of the BTW-FES dynamics and the structure of the network leads to the emergence of a bistable region between the
absorbing and active states.

The sustained oscillations also appear in some cyclically three-state systems such as the susceptible-infected-recovered-susceptible epidemic model \cite{SIRS} and the rock-paper-scissor game \cite{RPS}. For these systems, a phase transition occurs from an absorbing state to a self-sustained oscillation state if the network randomness exceeds a threshold value. Hence, the small-world effect results in qualitatively similar behaviors between the BTW-FES model with a symmetrical toppling rule and the cyclic three-state models. We propose the BTW-FES model has an inherent cyclic three-state dynamics that results in periodic oscillations. However finding an exact equivalence of the FES models and the cyclic three-state models is a challenge for future work.


\begin{thebibliography}{99}

\bibitem{Bak}
P. Bak, C. Tang, K. Wiesenfeld, Phys. Rev. Lett. {\bf 59}, 381 (1987).


\bibitem{Tang}
C. Tang, P. Bak, Phys. Rev. Lett. {\bf 60}, 2347 (1988).


\bibitem{wat}
N. W. Watkins, G. Pruessner, S. C. Chapman, N. B. Crosby, and H. J. Jensen, Space Sci. Rev. {\bf 198}, 3 (2016).


\bibitem{Dhar}
D. Dhar, Phys. Rev. Lett. {\bf 64}, 1613 (1990); {\bf 64}, 2837(E) (1990).

\bibitem{Dhar2}
S. N. Majumdar, D. Dhar, J. Phys. A: Math. Gen. {\bf 24}, L357 (1991).

%
%
%
%
%
%

\bibitem{Dikman}
R. Dickman, A. Vespignani, S. Zapperi, Phys. Rev. E {\bf 57}, 5095 (1998).


\bibitem{FES}
A. Vespignani, R. Dickman, M. A. Mu\~noz , S. Zapperi, Phys. Rev. E {\bf 62}, 4564 (2000).

\bibitem{FES2}
A. Vespignani, R. Dickman, M. A. Mu\~noz, S. Zapperi. Phys. Rev. Lett. {\bf 81}, 5676 (1998).

\bibitem{Bagnoli}
F. Bagnoli, F. Cecconi, A. Flammini, A. Vespignani, Europhys. Lett. {\bf 63}, 512 (2003).

\bibitem{Lucca}
L. Dall'Asta, Phys. Rev. Lett. {\bf 96}, 058003 (2006).


\bibitem{PT1}
V. Sidoravicius, A. Teixeira,  Electron. J. Probab. {\bf 22}, 1 (2017).

\bibitem{PT2}
R. Dickman, L. T. Rolla, V. Sidoravicius,  J. Stat. Phys. {\bf 138}, 126 (2010).


\bibitem{montakhab}
A. Montakhab and J. M. Carlson, Phys. Rev. E {\bf 58}, 5608 (1998).

\bibitem{path}
R. Dickman, M. A. Mu\~ noz, A. Vespignani, S. Zapperi, Brazilian Journal of Physics {\bf 30}, 27 (2000).


\bibitem{munoz1}
S. di Santo,  R. Burioni, A. Vezzani, M. A. Mu\~ noz, Phys. Rev. Lett. {\bf 116}, 240601 (2016).

\bibitem{munoz2}
V. Buend\'ia, S. di Santo, J. A. Bonachela, M. A. Mu\~ noz, Front. Phys. {\bf 8}, 333 (2020).


\bibitem{Bonabeau}
E. Bonabeau, J. Phys. Soc. Jpn. {\bf 64}, 327 (1995).


\bibitem{Goh1}
K. I. Goh, D. S. Lee, B. Kahng, D. Kim, Phys. Rev. Lett. {\bf 91}, 148701 (2003).


\bibitem{Goh2}
D. S. Lee, K. I. Goh, B. Kahng, and D. Kim, Physica A {\bf 338}, 84 (2004).


\bibitem{adaptive1}
T. Gross and H. Sayama, editors, \textit{Adaptive Networks: Theory, Models and Applications} (Springer-Verlag Berlin Heidelberg, 2009).


\bibitem{adaptive2}
T. Gross, C. J. Dommar D\'Lima, B. Blasius, Phys. Rev. Lett. {\bf 96}, 208701 (2006).


\bibitem{Bianconi}
G. Bianconi, M. Marsili, Phys. Rev. E {\bf 70}, 035105(R) (2004).


\bibitem{coevolution}
P. Fronczak, A. Fronczak, J. A. Holyst, Phys. Rev. E {\bf 73}, 046117 (2006).


\bibitem{watts}
D. J. Watts, S. H. Strogatz, Nature {\bf 393}, 440 (1998).


\bibitem{SIRS}
M. Kuperman, G. Abramson, Phys. Rev. Lett. {\bf 86}, 2909 (2001).


\bibitem{RPS}
G. Szab\'o, A. Szolnoki, and R. Izs\'ak, J. Phys. A {\bf 37}, 2599 (2004).



\end{thebibliography}
\end{document}